\documentclass[aps,twocolumn,amsmath,amssymb,prb,showpacs]{revtex4}
\usepackage[dvips]{graphicx}

\begin{document}

\title{Localization of Wannier functions for entangled energy bands}

\author{U.~Birkenheuer}
\thanks{Corresponding author}
\email{birken@mpipks-dresden.mpg.de}
\author{D.~Izotov}
\email{izotov@mpipks-dresden.mpg.de}%
\affiliation{Max-Planck-Institut f{\"u}r Physik komplexer Systeme,
             N{\"o}thnitzer Str.~38, 01187 Dresden, Germany}

\begin{abstract}
A new method for the localization of crystalline orbitals for
entangled energy bands is proposed. It is an extension of the
Wannier-Boys algorithm [C.M.~Zicovich-Wilson, R.~Dovesi, and
V.R.~Saunders, J.~Chem.~Phys.~{\bf 115}, 9708 (2001)] which
is particularly well-suited for 
linear combination of atomic orbital representations of the
Bloch waves. It allows the inclusion of additional bands during
the optimization of the unitary hybridization matrix used in 
the multi-band Wannier transformation. By a projection technique,
the proper chemical character is extracted from the Bloch waves and
compact localized orbitals are obtained even for entangled bands.
The performance of the new projective Wannier-Boys localization is 
demonstrated on the low-lying unoccupied bands of
{\it trans}-polyacetylene, diamond and silicon.
\end{abstract}

\pacs{71.15.Ap,71.15.Dx}

\maketitle

\section{Introduction}

Localization of Wannier functions (WF) has attracted great
attention of scientists in the recent years. Despite the success of
describing most of the physical phenomena in crystals in
terms of Bloch waves (BW), Wannier functions have obvious advantages. To
mention a few: WFs provide a chemically intuitive picture of the 
electronic structure in crystals, using localized WFs, physical
quantities such as effective Born charges and spontaneous polarization
can be evaluated in a very simple way,~\cite{King-Smith93,Resta93,Resta94}
and they play a central role in many post Hartree-Fock electron correlation
methods.~\cite{Stoll92si,Stoll92c,Graefenstein93,Graefenstein97,Albrecht00}

Several rigorous schemes for the localization of WFs have been
proposed so far. They fall into two categories, those which are
based on the variational principle,~\cite{Kohn73,Smirnov01}
and those which are an extension of the Foster-Boys localization
criterion \cite{FosterBoys1,FosterBoys2} for periodic
systems.~\cite{Marzari97,Berghold00,Baranek01,Zicovich01} All
these methods can only be applied to isolated band complexes,
i.e.\ a group of bands which are separated from the other 
bands by an energy gap over the {\it entire} Brillouin zone.
This restriction appreciably confines the possible
applications of those methods and only a limited number of systems
can be treated.
In particular, the energy bands of the unoccupied Bloch waves usually
do not exhibit any such band gaps. Thus, none of the localization
schemes developed so far can be applied routinely to selectively generate
virtual Wannier functions as needed, for instances, in our wave-function-based
post Hartree-Fock correlation methods for valence {\it and} conduction
bands.~\cite{Bezugly04,Willnauer04}

Recently, Souza {\it et al}.~\cite{Souza01} have extended the original
Marzari-Vanderbilt localization scheme~\cite{Marzari97} to systems with
entangled bands. The method is based on a preselection of optimal Bloch waves
having maximal similarity at neighboring $k$-points by minimizing a suitable
functional. Like the original scheme, the new method
is especially designed for a plane wave representation of the BWs and
heavily relies on numerical $k$-space differentiations.

In this paper, we propose a similar extension for the Wannier-Boys
localization scheme developed by Zicovich-Wilson, Dovesi and 
Saunders.\cite{Zicovich01} This scheme differs in many aspects from the
algorithm proposed by Marzari {\it et al}.~\cite{Marzari97} In particular,
it is much better suited for BWs given in a linear combination of atomic-like
orbital (LCAO) representation as employed in many widely used program packages
for periodic systems such as {\sc Crystal},~\cite{CRYSTAL03}
{\sc Gaussian},~\cite{GAUSSIAN03} {\sc DMol}$^3$,~\cite{DMOL3}
{\sc Nfp-Lmto},~\cite{LMTOxx} or {\sc Band}~\cite{BAND04} (the periodic
variant of {\sc Adf}\cite{ADF04}).

In Sec.~\ref{sec:local}, the details of our algorithm are described.
Then, in Sec.~\ref{sec:res}, the results from the localization 
are presented for {\it trans\/}-polyacetylene (tPA),
diamond and silicon, and some concluding remarks are drawn in
Sec.~\ref{sec:concl}.

\section{The Localization procedure}\label{sec:local}
\newcommand{\rrvec}{{\mbox{\normalsize \boldmath $r$}}}
\newcommand{\ssvec}{{\mbox{\normalsize \boldmath $s$}}}
\newcommand{\kkvec}{{\mbox{\normalsize \boldmath $k$}}}
\newcommand{\RRvec}{{\mbox{\normalsize \boldmath $R$}}}
\newcommand{\rvec }{{\mbox{\scriptsize \boldmath $r$}}}
\newcommand{\svec }{{\mbox{\scriptsize \boldmath $s$}}}
\newcommand{\kvec }{{\mbox{\scriptsize \boldmath $k$}}}
\newcommand{\Rvec }{{\mbox{\scriptsize \boldmath $R$}}}

The new projective Wannier-Boys scheme we want to present here sets
out from the original Wannier-Boys (WB) localization procedure which is
discussed in detail in Ref.~\onlinecite{Zicovich01}.
Like all localization schemes for composite bands, it relies on the initial
specification of a fixed set of energy bands. These bands determine the space
of the Bloch functions which are allowed to participate in the multi-band
Wannier transformation, the so-called active space. For example,
the valence bands of a system can be chosen as such a set of bands. The WB
algorithm is a combination of two steps: the so-called 'wannierization' and
a Foster-Boys localization of the obtained WFs within the reference unit cell.
Recently, the algorithm has been extended to operate with a multi-cell 
Foster-Boys localization to better preserve the space group symmetry
of the system under consideration.~\cite{Zicovichxx} 

The wannierization step starts from a set of
trial Wannier functions $\omega_s^{(0)}(\rrvec)$ which
are linear combinations of atomic-like orbitals
\begin{equation}
  \omega_s^{(0)}(\rrvec) = \sum_{\mu,\Rvec} c^{\Rvec}_{\mu s}
                           \phi_\mu(\rrvec-\ssvec_\mu-\RRvec)
  \quad.
\end{equation}
We follow the notation from Ref.~\onlinecite{Zicovich01} here. Thus, 
$\mu$ runs over all atomic basis functions
$\phi_\mu(\rrvec-\ssvec_\mu)$ in the reference
unit cell, $\ssvec_\mu$ denotes their centers, and $\RRvec$ runs over
all lattice vectors of the underlying Bravais lattice.
To reduce the spacial extent of each of these WFs, the orbital
coefficients $c^{\Rvec}_{\mu s}$ are set to zero for all sites
$\ssvec_\mu+\RRvec$ at which the atomic Mulliken
populations\cite{Mulliken1,Mulliken2}
of the given Wannier function falls below a certain threshold (for
details see Ref.~\onlinecite{Zicovich01}). The WFs obtained this way,
the so-called `model functions',\cite{Zicovich01}
are transformed to $k$-space, projected onto the active space
spanned by the selected BWs, orthonormalized again, transformed back to
real space, and moved back into the reference unit cell (if necessary). 
The resulting (real) WFs, $\overline{\omega}^{(1)}_s$, then enter the 
Foster-Boys step, where they are subject to an orthogonal transformation
\begin{equation}
  \omega_s^{(1)} = \sum_{t=1}^N \overline{\omega}_t^{(1)} O_{ts}
\end{equation}
that minimizes the spread $\Omega[\{\omega_s^{(1)}\}]$ given by
the functional
\begin{equation}\label{eq:FB}
  \Omega[\{\omega_s\}] = \sum_{s=1}^N\Bigl(\langle\omega_s |r^2| \omega_s\rangle
                       - \langle\omega_s |\rrvec| \omega_s\rangle^2\Bigr)
  \quad.
\end{equation}
Here, $N$ is the number of energy bands involved in the localization.
Finally, the optimized functions $\omega_s^{(1)}$ are used as new (orthonormal)
trial functions for the wannierization and the whole procedure is repeated until
convergence is reached. The discarding of orbital coefficients and the
subsequent projection onto the active space is the crucial part of the WB
algorithm. It is combined with a Foster-Boys localization to ensure
localization of the WFs also {\it inside} the unit cells.

The described algorithm performs well for isolated band complexes. In the
case of entangled bands, however, the selection of proper bands to set up
a suitable active space becomes problematic. Avoided and symmetry allowed
crossings between the energy bands in mind and other, disturbing energy bands
occur, and 
the orbital character one is looking for is spread over several BWs
which in turn exhibit more or less strong admixtures from other,
contaminating orbitals. To overcome this difficulties give up the 
concept of an rigid active space, abandon the 
constraint that the number of selected Bloch waves per $k$-point has to
coincide with the number $N$ of Wannier functions per unit cell, and allow
additional BWs to be included in the active space at each $k$-point.

The selection of an appropriate $N$-dimensional active subspace is then done
in the projection step during the wannierization. To this end the `model Bloch
waves'
\begin{equation}
  \xi_{s\kvec}(\rrvec) = \sum_{\Rvec} {\rm e}^{{\rm i}\kvec\Rvec}
                         \xi_s(\rrvec-\RRvec)
  \quad,
\end{equation}
which are the $k$-space transformed of the model functions $\xi_s(\rrvec)$,
are projected onto the active space via
\begin{equation}\label{eq:proj}
  \xi^{\prime}_{s\kvec} = \sum_{n=1}^{N_{\kvec}} \psi_{n\kvec}
                          \langle \psi_{n\kvec} | \xi_{s\kvec} \rangle
  \quad\mbox{for } s = 1,\ldots\,N
\end{equation}
with the number $N_{\kvec}$ of selected BWs $\psi_{n\kvec}$ at each $k$-point
being at least as large as the number of WFs per unit cell. The
matrices $U_{ns}(\kkvec) = \langle \psi_{n\kvec} | \xi_{s\kvec} \rangle$
showing up here establish a generalization of the unitary hybridization matrices
used in conventional multi-band Wannier transformations.\cite{multiWT} The new
projected functions $\xi^{\prime}_{s\kvec}$ are those functions in the
active space which resemble the initial model Bloch waves $\xi_{s\kvec}$ the
most. They span the $N$-dimensional subspace used in  
the subsequent multi-band Wannier transformation. In this sense, the
procedure outlined here is very similar to the one proposed by Souza
{\it et al}.~\cite{Souza01} In particular, our extended projection step during
the wannerization can be regarded as an analog to the band preselection
scheme used in their method.

As for any hybrid orbitals, the orbital energies
\begin{equation}
  \varepsilon_{s\kvec} = \langle \xi^{\prime}_{s\kvec} | F |
                                 \xi^{\prime}_{s\kvec} \rangle
\end{equation}
of the projected model Bloch waves have little in common with the canonical
band energies $\varepsilon_{n\kvec}$ they originate from. Yet, by
diagonalizing the subblock
\begin{equation}\label{eq:Fst}
  F_{st}(\kkvec) = \langle \xi^{\prime}_{s\kvec} | F |
                           \xi^{\prime}_{t\kvec} \rangle
\end{equation}
of the Fock operator $F$ of the
system, $N$ new, so-called disentangled energy bands $\eta_{s\kvec}$ can be
obtained. Where the contamination of the canonical BWs with orbitals of wrong
character is small, the disentangled bands will essentially coincide with the
canonical ones. Close to band crossings, where the contamination is larger,
they will deviate substantially from the canonical bands in order to be
able to follow the chemical
nature of the underlying BWs. By this the disentangled bands form 
an {\it effectively} isolated complex 
of $N$ bands with none of them showing any kinks and cones (see
Sec.~\ref{sec:res} for more details). The eigenvectors resulting from the
diagonalization of $F_{st}(\kkvec)$ can be regarded as sort of optimal Bloch
wave hybrids with
minimal orbital contamination and vanishing off-diagonal terms in the Fock
operator. We will refer to these hybrids as disentangled Bloch waves.

The {\it canonical} Bloch waves $\psi_{n\kvec}$, to be included in the active
space, can be selected in various ways, for example, by specifying an energy
window and taking all BWs whose band energies $\varepsilon_{n\kvec}$ fall into
this window. Alternatively, a so-called 'energy tube' around a given pair
of reference bands $(\underline{n},\overline{n})$ may be used, i.e. all BWs
with band energies
\begin{equation}
  \varepsilon_{n\kvec} \in
  [ \varepsilon_{\underline{n}\kvec} - \underline{\varepsilon} ,
    \varepsilon_{\overline {n}\kvec} + \overline {\varepsilon} ]
\end{equation}
are considered where $\underline{\varepsilon}$ and $\overline{\varepsilon}$
are some user-specified energy tolerances.

We have implemented the above projection and rediagonalization scheme
as an extension to the original WB localization routine in the 
{\sc Crystal} 200x code,~\cite{CRYSTAL00} a precursor of the most recent public
version of the {\sc Crystal} program package.~\cite{CRYSTAL03}
Its ability to disentangle energy bands properly will be demonstrated 
in Sec.~\ref{sec:res} where our method is applied to the virtual
bands of {\it trans}-polyacetylene, diamond and silicon.

\section{Results and discussion}\label{sec:res}

All band structures shown here are calculated on the
Hartree-Fock level of theory. The periodic {\it ab initio\/}
program package {\sc Crystal} (version 200x)~\cite{CRYSTAL00} is used for 
that purpose. The localization of the WFs is performed
a posteriori with the WB algorithm \cite{Zicovich01} as implemented in
{\sc Crystal} 200x\cite{CRYSTAL00}
in conjunction with our extension for entangled bands which has been
built into this version of {\sc Crystal}.

In all cases, we focus on the first few, low-lying virtual bands
of our systems. Because of the larger extent of the 
localized virtual Wannier functions compared to the occupied ones,
the former WFs are quite sensitive to the number of $k$-points in
the Monkhorst-Pack grid.\cite{Monkhorst-Pack} We chosed the grids sufficiently
fine to remove any ambiguities resulting from the discrete $k$-space
integration.

\subsection{trans-Polyacetylene}\label{sec:tPA}
\begin{figure}[htb]
\centering
\includegraphics[width=0.35\textwidth]{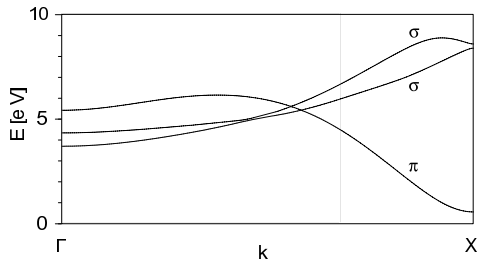}\\%
\caption{Virtual Hartree-Fock bands of {\it trans}-polyacetylene in the energy
         range from 0 to 10~eV calculated with a cc-pVTZ basis
         set.\cite{Dunning89}}
\label{tPA:bands}
\end{figure}

{\it trans}-Polyacetylene (tPA), --[HC=CH]$_\infty$--, suits perfectly as
illustrative example for band
disentanglement, because, in the basis set employed here, it exhibits three
low-lying entangled virtual bands which are separated from the rest of
the unoccupied band structure (Fig.~\ref{tPA:bands}).

The computational parameter for tPA are taken from a recent study on the
correlated valence and conduction bands of tPA.\cite{Bezugly04} That is,
Dunning's correlation-consistent polarized valence triple-$\zeta$ basis sets
(cc-pVTZ)~\cite{Dunning89} for hydrogen and carbon (without $f$ functions)
are used, the bond distances are d(C-C) = 1.45~\AA, d(C=C) = 1.36~\AA, and
d(C-H) = 1.09~\AA, and the bond angles are $\angle$(C-C=C) = 121.7$^\circ$ and
$\angle$(C-C-H) = 118.2$^\circ$. Two different geometries are considered, 
flat tPA (the experimental structure) and a distorted configuration with the
hydrogens being bent out-of-plane by 20$^\circ$ in such a way that the
inversion symmetry of the polymer is preserved. The Brillioun zone is sampled
by a uniform grid of 100 $k$-points.

The first three virtual bands are selected for the disentanglement
(Fig.~\ref{tPA:bands}). One is
a $\pi^\ast$ band formed by C=C $\pi$ anti bonds, the other two are 
of $\sigma$ symmetry and describe $\mbox{C--H}$ anti bonds (not $\mbox{C--C}$
$\sigma$ anti bonds, as one might think at first glance). For the flat polymer
the symmetry separation is perfect, for the distorted structure some
mixture between $\sigma$ and $\pi$ bonds occurs. Nevertheless, the 
two types of BWs remain quite different in their orbital character 
which should facilitate the band disentanglement significantly.
In this sense, our first system very much resembles
the one chosen by Souza {\it et al.}~\cite{Souza01} They used copper which 
exhibits a $d$ band manifold which is entangled with a single $sp$ valence
band. 

Of course, for tPA one could 
localize the virtual Bloch waves by means of the original WB algorithm. But
what we want to demonstrate here, is that it is also possible to localize the
$\pi$ and $\sigma$ bands {\it separately}. 
We first consider the flat tPA chain. In that case, the BWs come in two 
different symmetries and the disentanglement
could simply be achieved by a proper labeling of the energy bands and the
associated Bloch waves. Yet, such symmetry classifications are hard to 
implement in localization schemes for periodic systems, and thus usually
not exploited. Our band disentanglement algorithm, however, is able to
recognize the different symmetries and to separate the bands properly.

As an initial guess for the $\pi$-type WFs, anti-phase linear combinations of  
$2p_z$ atomic orbitals at neighboring carbon atoms are used. For
the $\sigma$-type WFs, anti-bonding linear combinations of $2sp^2$ hybrid
orbitals on carbon and $1s$ atomic orbitals on hydrogen are constructed.
Because of this choice, the hybridization matrices $U_{ns}(\kkvec)$ become
$3\times 1$ and $3\times 2$ matrices, respectively, with a maximum-rank
subblock and all other entries being exactly zero, as is confirmed numerically.

\begin{figure}[htb]
\centering
\includegraphics[width=0.35\textwidth]{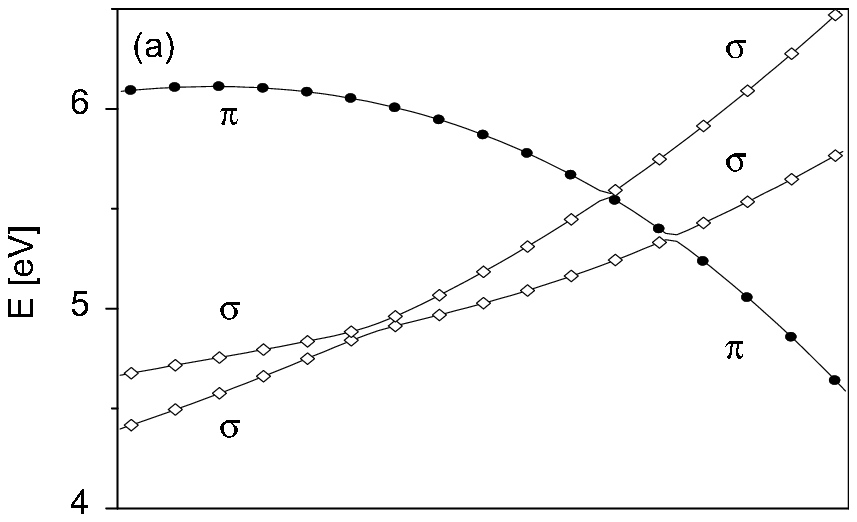}\\%
\includegraphics[width=0.35\textwidth]{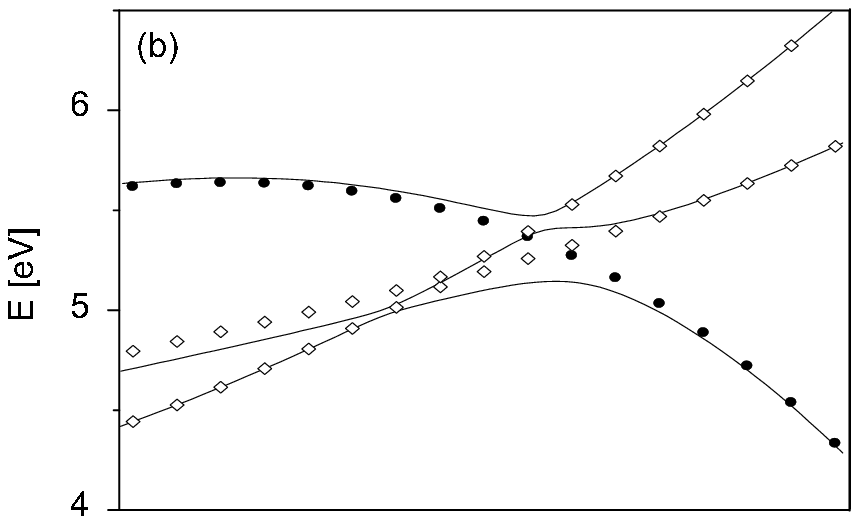}\\%
\caption{Zoom into the first three virtual bands of {\it trans}-polyacetylene
         along the $\Delta$ line (the endpoints of the plot correspond to 1/3
         and 2/3 of the $\Gamma$X distance) for the flat (a) and the distorted
         (b) geometry of the polymer. Solid lines show the canonical bands,
         symbols represent the disentangled bands: one of $\pi^\ast$ character
         ($\bullet$), two of $\sigma^\ast$ character ($\diamond$).}
\label{fig:tPA}
\end{figure}

As seen in the upper panel of Fig.~\ref{fig:tPA}, where the canonical band
energies $\varepsilon_{n\kvec}$ are compared to the disentangled band
energies $\eta_{s\kvec}$, our band disentanglement procedure is perfectly
able to describe either the $\pi$ or the $\sigma$ orbitals alone.
The crossing of the $\sigma$ bands on the left-hand side is an avoided one
(because of the lack of any extra symmetry in the small point group of the
$k$-points inside the Brillouin zone). The apparently extremely weak coupling
of the two bands is due to `soft symmetry selection rules'.\cite{softsym}
That means, the impact of the C-C bond length alternation in tPA on the C--H
anti-bond dominated $\sigma^\ast$ bands is so small that they essentially 
behave as if there would exist an additional glide plane symmetry in the system
(like in equidistant tPA). This concept is corroborated by a detailed analysis
of the involved Bloch waves.

A more interesting situation for band disentanglement arises when the
symmetry of the system is lowered by moving the hydrogen atoms out of plane.
Now, the three energy bands avoid each other and the underlying Bloch waves
carry both, $\pi$- and $\sigma$-type atomic orbital contributions.
Using band disentanglement we are able to follow this contributions
individually. Setting out from an initial guess for $\pi$-type WFs, a single
smooth band can be generated starting at the third canonical band at the 
$\Gamma$ point, passing the avoided crossing in the middle of the Brillouin zone
without any kinks and wiggles and ending at the lowest band at the X point
(see black dots in the Fig.~\ref{fig:tPA}b).
The same holds for the two disentangled band of predominantly $\sigma$ 
character. When going from $\Gamma$ to X they smoothly switch from the lower
two bands to the upper two bands without being influenced by the complex
structure of the canonical bands close to the multiple avoided crossing.

After having demonstrated that our projective Wannier-Boys scheme is 
able to separate energy bands appropriately, the effect of the 
disentanglement on the locality of the resulting Wannier functions
should be addressed. For that purpose we turn our attention to the 
more complex case of bulk materials, diamond and silicon in our case.

\subsection{Diamond}
Because of the rather diffuse, atom-optimized basis functions present in the
original 
carbon cc-pVTZ basis set of Dunning,\cite{Dunning89} (outermost $s$ and $p$
exponents of 0.1285 and 0.1209, respectively) it can not be used
for a Hartree-Fock calculation of bulk diamond. Hence, the 
outermost exponents were reoptimized
by minimizing the Hartree-Fock energy per unit cell of diamond. The resulting
exponents are 0.2011 for the $s$ function and 0.6256, 0.3243 for the $p$
functions, typical values for diamond.\cite{Cbas-Torino}
In addition the two $d$ 
functions of the triple-$\zeta$ basis set were replaced by the single one
of the corresponding double-$\zeta$ basis set
(with exponent 0.55),\cite{Dunning89} and, as for tPA, the $f$ function 
had to be skipped because {\sc Crystal} cannot handle them.
This basis set, referred to
as bulk-optimized cc-pVTZ, has been used very successfully in our embedding 
studies of wave-function-based correlation calculations for
diamond.\cite{Willnauer04} The experimental lattice constant of
3.57~\AA\cite{alat} is adopted which corresponds to an interatomic C-C
distance of 1.546~\AA, together with a $40\times 40\times 40$ Monkhorst-Pack
grid.

\begin{figure}[htb]
\centering
\includegraphics[width=0.4\textwidth]{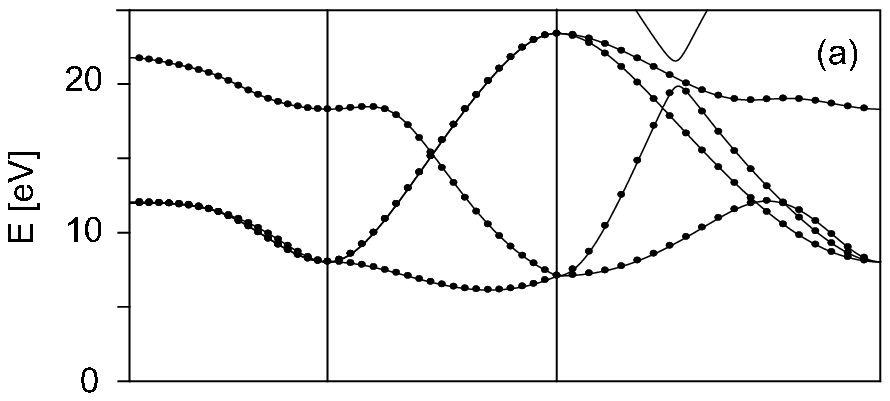}\\%
\includegraphics[width=0.4\textwidth]{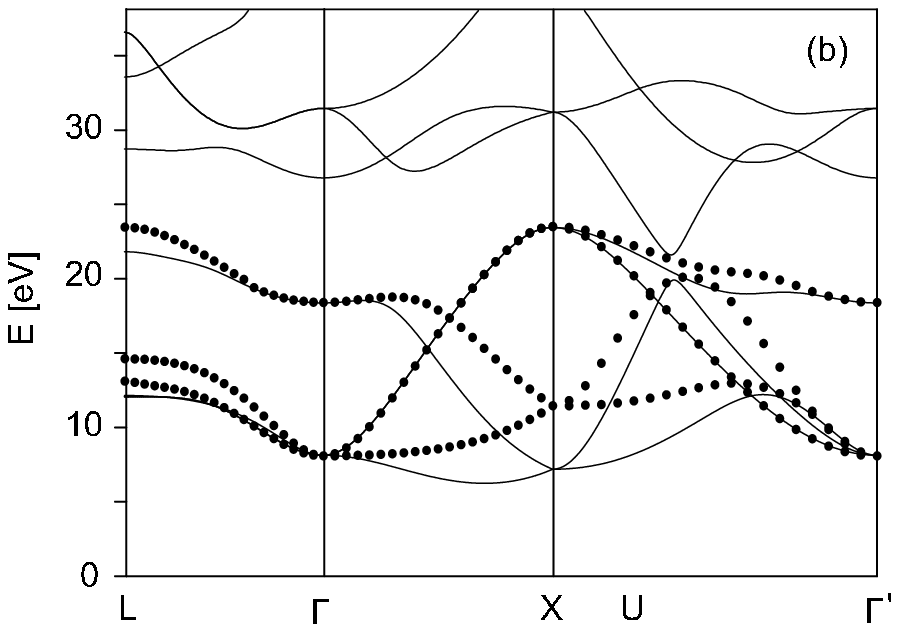}\\%
\caption{Virtual band structure of diamond from a canonical Hartree-Fock
         calculation (solid lines) in comparison to the band energies
         resulting from localized Wannier functions (dotted lines).
         The localization is either done with the original WB algorithm (a)
         or by band disentanglement with an energy window of $0-38$~eV (b).}
\label{fig:c_bands}
\end{figure}

\begin{figure}[htb]
\centering
\includegraphics[width=0.3\textwidth]{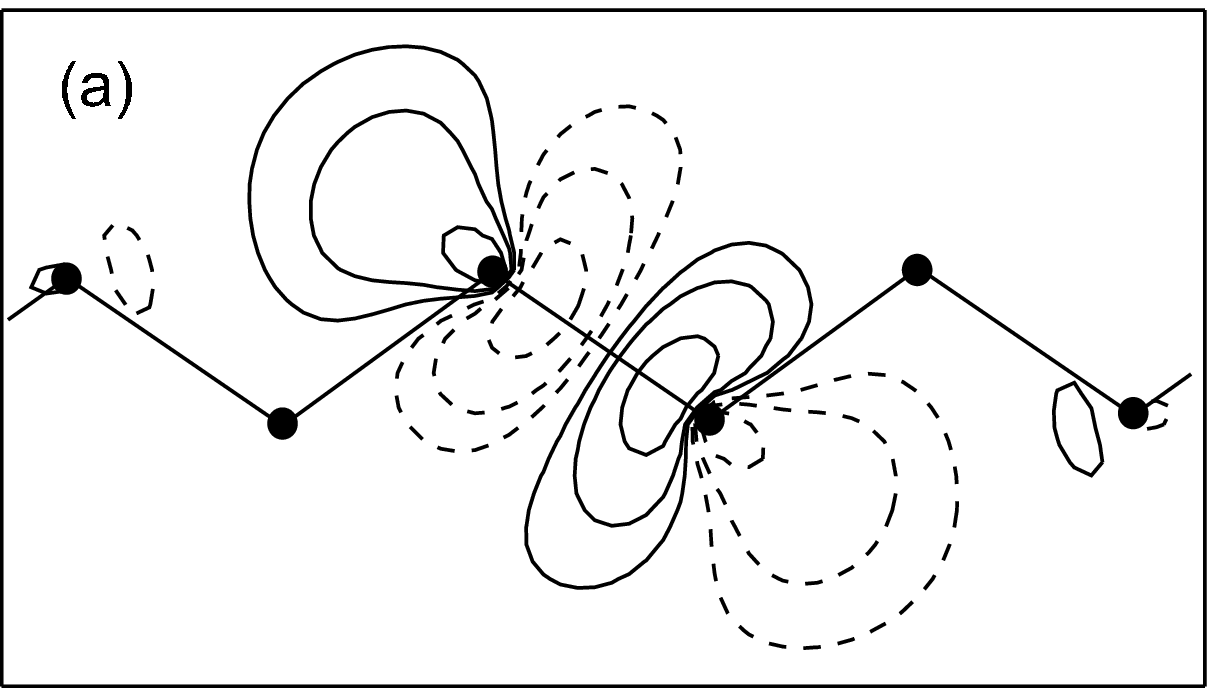}\\%
\includegraphics[width=0.3\textwidth]{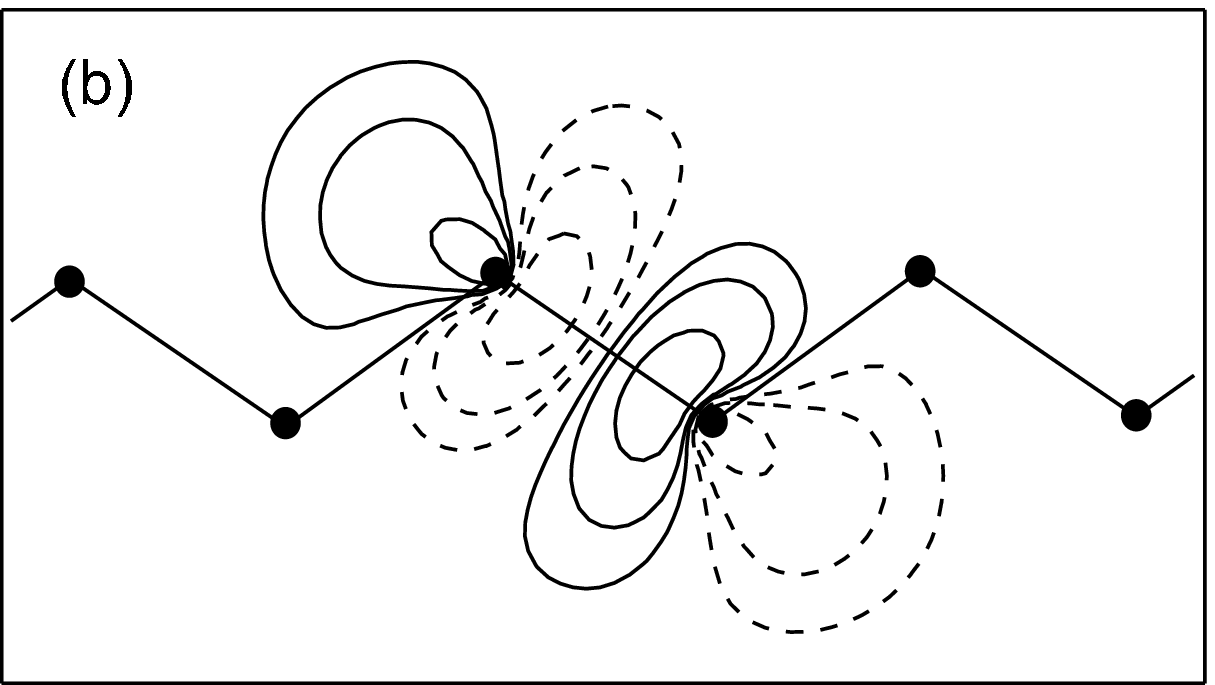}\\%
\caption{Contour plots of the virtual Wannier function of diamond shown in the 
         (110) plane of a C--C zig-zag chain as obtained
         by the original WB algorithm (a) or by band disentanglement with 
         an energy window of $0-38$~eV (b). The values of the contours are
         $\pm$0.046, $\pm$0.10 and $\pm$0.22 bohr$^{-3/2}$ (a geometrical
         progression with $q = 10^{1/3}$).}
\label{fig:lwo_c}
\end{figure}

Because the first four low-lying virtual bands of diamond are separated from
the rest of the virtual bands by a small gap, it is possible to use
the original WB algorithm to generate localized Wannier functions $\omega_s$.
The energy bands obtained from the Wannier representation
$F_{st}(\RRvec) = \langle\omega_s(\rrvec)|F|\omega_t(\rrvec-\RRvec)\rangle$
of the Fock operator --- or equivalently from $F_{st}(\kkvec)$ as defined in
Eq.~(\ref{eq:Fst}) --- exactly reproduces the canonical bands
(see Fig.~\ref{fig:c_bands}a). Nevertheless, the resulting localized Wannier
functions possess rather substantial tails at the second nearest
neighbor carbon atoms as seen in Fig.~\ref{fig:lwo_c}a.
These tails can very-well spoil the performance of any scheme
which relies on the locality of virtual WFs (like the one having been used in
Ref.~\onlinecite{Bezugly04}).

Extending the active space used in the WB algorithm opens the possibility for
making the Wannier
functions more compact. Switching to an energy window of $0-38$~eV the spread
of the virtual Wannier function, as measured by the Foster-Boys functional
(\ref{eq:FB}), reduces by more than a factor of two
(see table~\ref{tab:spread}) and the undesired
orbital contributions at the second nearest neighbor carbon
atoms disappear in the contour plot shown in Fig.~\ref{fig:lwo_c}b.
No shrinking of the virtual WF around the central bond is
discernible in Fig.~\ref{fig:lwo_c} compared to the WF obtained by the
original WB localization. Apparently, the reduction in the spread is 
solely due to the fading of the tails in the localized Wannier
function, precisely what we are aiming for.

The price to pay, is an overall upwards shift of the disentangled bands 
with respect to the canonical ones, more pronounced at the X point than 
at the $\Gamma$ or L point, a phenomenon we also hit on for
silicon in Sec.~\ref{sec:si}. It is a feature one often observes when 
Foster-Boys-like schemes are used to localize virtual orbitals. The 
localization functional $\Omega$ tries to minimize the extent of the 
orbitals as much as possible, regardless of the chemical nature of the 
orbitals and, in particular, their orbital energies. But compactness of
orbitals usually implies high kinetic energies. Thus, it can easily happen
that the WFs pick up more and more kinetic energy during the iterative
procedure of the projective WB localization algorithm as soon as the BWs
spanning active space allow so. In practice, a compromise has to be found
between tracing the proper orbital character in BWs energetically far away
from the bands in mind and the risk of opening channels for spurious
orbitals compression.

\begin{table}[htb]
\caption{The spread $\Omega$ (per orbital) of the virtual Wannier functions of
         diamond and silicon (in bohr$^2$) as a function of the energy window 
         (in eV) used for the band disentanglement. For comparison, the scaled
         spreads $\Omega/a^2$ with $a$ being the lattice constant are given as
         well.}
\label{tab:spread}%
\begin{ruledtabular}
\begin{tabular}{lrrlrr}
\multicolumn{3}{c}{diamond} & \multicolumn{3}{c}{silicon}\\
window & $\Omega$ & $\Omega/a^2$ & window & $\Omega$ & $\Omega/a^2$ \\
\hline
none   &    10.10 &        0.792 & $0-15$ &    33.03 & 1.120        \\
$0-38$ &     4.94 &        0.388 & $0-31$ &    15.45 & 0.524        \\
\end{tabular}
\end{ruledtabular}
\end{table}

\subsection{Silicon}\label{sec:si}
Silicon is the last and most interesting example which is discussed here,
because local virtual Wannier functions can not be generated at all for
silicon without band disentanglement.

The relativistic energy-consistent Ne-core pseudopotential from
Stuttgart\cite{Bergner93} together with a decontracted $[3s3p]$ version
of the corresponding optimized valence double-$\zeta$ basis set are
used here. The basis set is augmented by a single $d$ polarization function
with an exponent of 0.4. The Si-Si distance is set to 2.352~{\AA} which
corresponds to a lattice constant of 5.432~{\AA}. These
computational parameters originate from the first pioneer study
on a rigorous determination of the correlation energy of silicon by means of
an incremental expansion,\cite{Stoll92si} and has successfully been used from
that time on in all {\it ab initio} study of correlation effects in bulk silicon
performed with the incremental scheme\cite{Paulus95,nonortho} or its extension
to valence and conduction bands.\cite{Graefenstein97,Albrecht00} Like for
diamond, a very dense $40 \times 40 \times 40$ Monkhorst-Pack grid is used here
to resolve the subtle details of the silicon conduction bands to be discussed
below.

\begin{figure}[htb]
\centering
\includegraphics[width=0.3\textwidth]{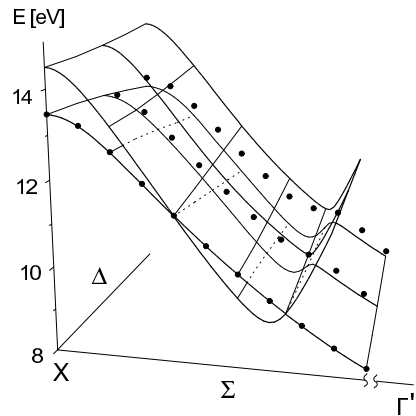}\\%
\caption{The fourth and fifth conduction band of silicon in the vicinity
         of the two symmetry-allowed crossings on the $\Sigma$ line drawn
         as a 2-dimensional function in the $\Gamma$X$\Gamma^\prime$ plane.
         The canonical bands are shown as energy surfaces, the
         uppermost disentangled band (whose energies are only
         available on the Monkhorst-Pack grid) as dots. An energy window
         of $0-15$~eV is used for the disentanglement.}
\label{fig:si_kiss}
\end{figure}

\begin{figure}[htb]
\centering
\includegraphics[width=0.4\textwidth]{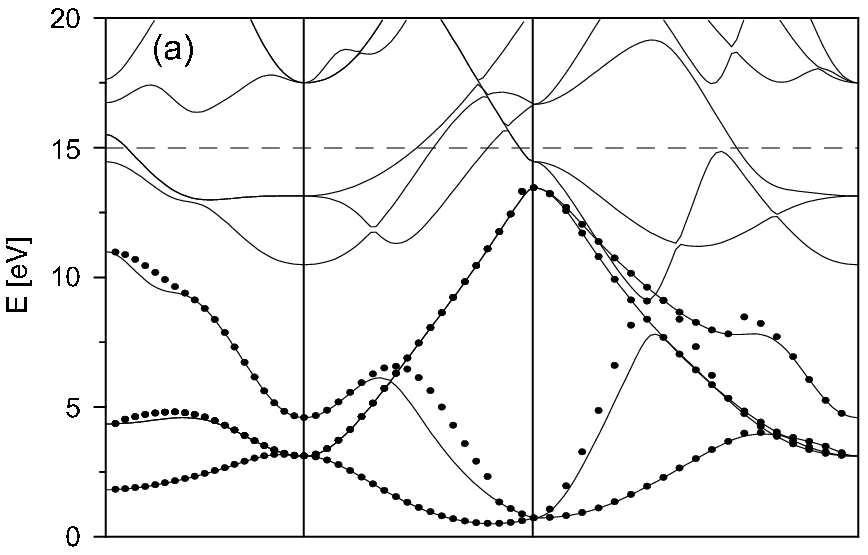}\\%
\includegraphics[width=0.4\textwidth]{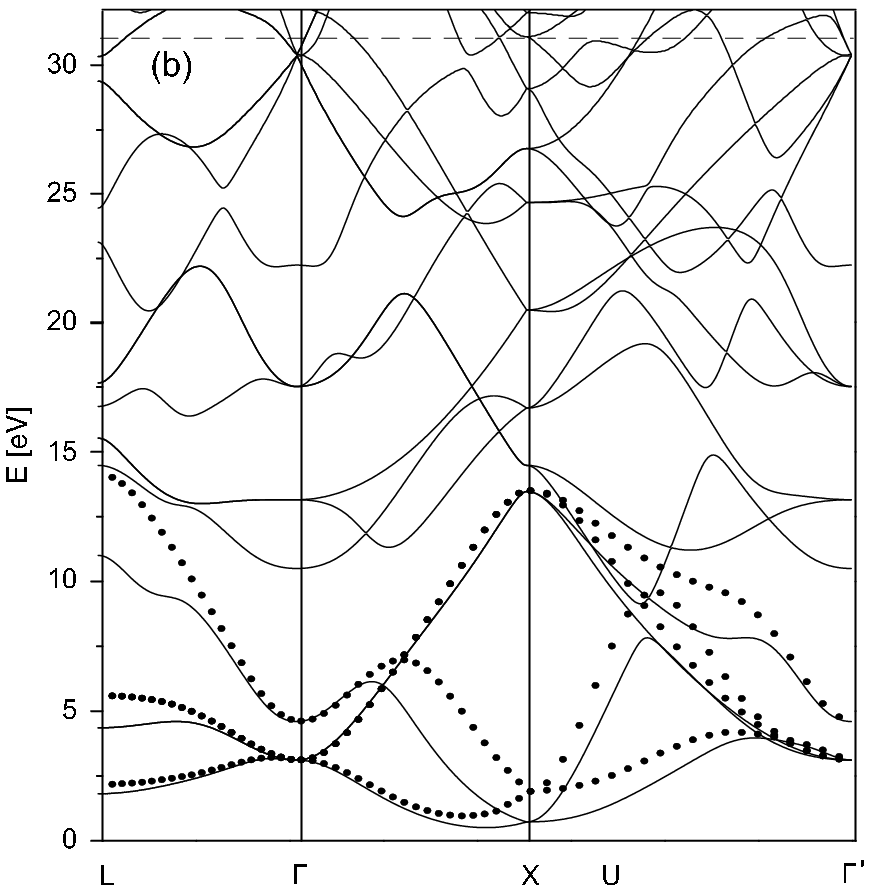}\\%
\caption{Virtual Hartree-Fock band structure of silicon (solid lines) and 
         the energies of the disentangled bands (dots) using either a 
         $0-15$~eV energy window (a) or a $0-31$~eV energy window (b)
         for the band disentanglement.}
\label{fig:si_bands}
\end{figure}

The reasons why band disentanglement is absolutely crucial for silicon
are the two symmetry-allowed crossing of a fourth and fifth conduction band
on the $\Sigma$ line from X over ${\rm U}={\rm K}$ to $\Gamma^\prime$ (the 
${\rm S}+\Sigma$ line
to be precise) which prevents a direct application of the WB
algorithm (see Figs.~\ref{fig:si_kiss} and \ref{fig:si_bands}). The
localization simply fails because the active BWs exhibit symmetries
different from the ones of the model Bloch waves $\xi_{s\kvec}$
and the projection step~(\ref{eq:proj}) yields linear dependent projections
$\xi^\prime_{s\kvec}$. A $d$ band is involved, one might speculate, but
closer inspection of the corresponding BWs reveals that the fifth conduction
band is an $sp$ band, formed --- in contrast to the other four conduction
bands --- by $s$ and $p$ orbitals of the next atomic shell following the
$3s/3p$ valence shell, at least in the basis set employed here.

Sometimes it is argued that the band crossing problem discussed above can
simply be solved by a proper relabeling of the energy bands. This is not the
case. Energy bands of bulk materials are three-dimensional functions of the
crystal momentum $\kkvec$ and the two critical bands
exhibit an interesting topology around the symmetry allowed
crossings. They only touch, twice, like the tips of two cones, as is clearly
seen in Fig.~\ref{fig:si_kiss} where the band energies are plotted as
2-dimensional energy surfaces over the $\Gamma$X$\Gamma^\prime$ plane.
The symmetry-allowed crossings are singularities. No band crossing occurs
anywhere else in the neighborhood of these points. Nevertheless, the character
of the associated Bloch waves still switches from one band to the
other when passing from the left-hand to the right-hand side of the plot plane
precisely as in conventional avoided crossings.

\begin{figure}[htb]
\centering
\includegraphics[width=0.3\textwidth]{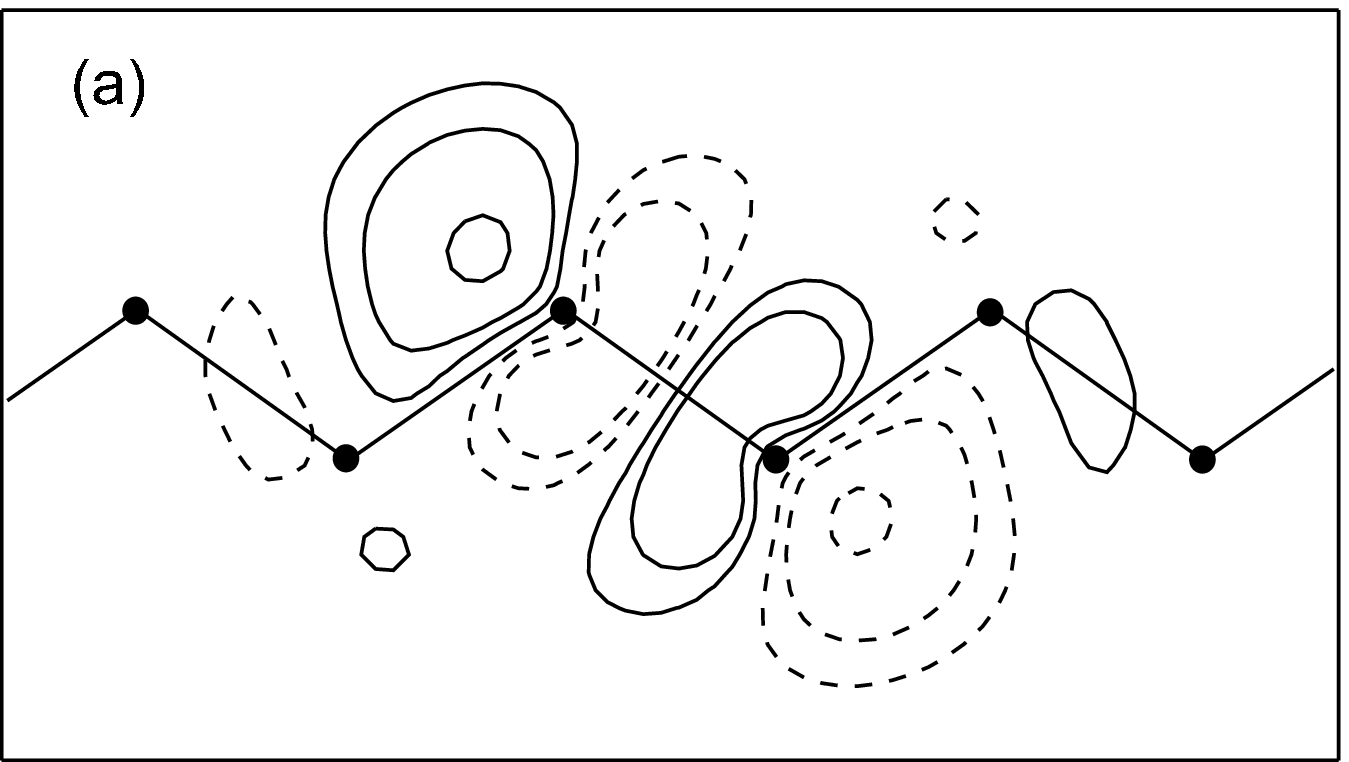}\\%
\includegraphics[width=0.3\textwidth]{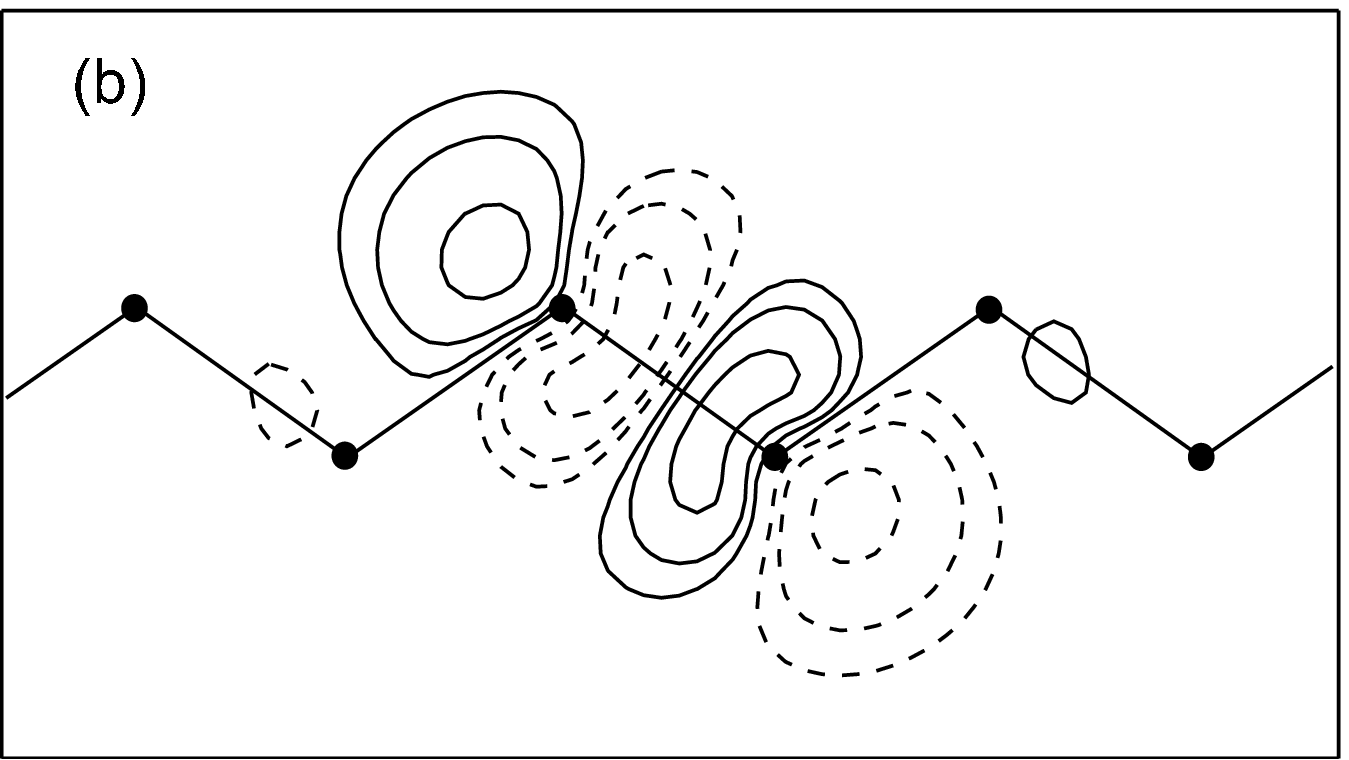}\\%
\caption{Contour plots of the virtual Wannier function in silicon shown in the 
         (110) plane of a Si--Si zig-zag chain as obtained
         from band disentanglement with an energy window of $0-15$~eV (a) or
         $0-31$~eV (b).
         The value of the contours are $\pm$0.022, $\pm$0.046 and $\pm$0.10
         bohr$^{-3/2}$ (see Fig.~\ref{fig:lwo_c}).}
\label{fig:si_lwo}
\end{figure}

Band disentanglement is the only way around. Two different energy windows are
considered, one being sort of minimal with the upper edge at 15~eV which is
closely above the top of the fourth conduction band, the other being big enough
to follow the BWs with proper $sp$ character up to the 12-th unoccupied
band at $\Gamma^\prime = \frac{2\pi}{a} (1,1,1)$.
In both cases, the localization could be performed without any problems.
As clearly seen in Fig.~\ref{fig:si_kiss}, the different symmetry of the
fourth and fifth conduction band along the $\Sigma$ line is perfectly 
recognized by the band disentanglement algorithm --- like for the flat tPA
chain discussed in Sec.~\ref{sec:tPA} --- and the uppermost disentangled energy
band exclusively follows the flatter of the two canonical bands without 
any kinks. Leaving the high-symmetry line, the disentangled bands start
to interpolate between the two canonical bands and form a smooth and
well-behaved energy surface.

The resulting localized virtual WFs of silicon are depicted in
Fig.~\ref{fig:si_lwo}. Compared to diamond which exhibits typical 
$sp^3$ hybrid character around the nuclei the virtual WF of silicon
is more symmetric and $p$-like in the vicinity of the nuclei. 
The same holds for the maximally localized virtual WF of silicon shown in
Ref.~\onlinecite{Souza01}.
Expectedly, the larger energy window yields the more compact WF, and
like for diamond, the reduction in the spread is impressive, from 33 down
to 15 bohr$^2$ (table~\ref{tab:spread}). Yet, all together, the 
virtual orbitals of silicon remain more diffuse than those of diamond
even if the difference in the lattice constants is accounted for, as
done by the scaled spreads listed in Table~\ref{tab:spread}.
This is not surprising, because of the much smaller (direct) band gap of
silicon (7.3~eV at $\Gamma$, experimentally\cite{gaps}) compared to diamond
(3.4~eV\cite{gaps}).

As already discussed above, the disentangled conduction bands have the
tendency to
shift upwards with increasing size of the energy window, a trend which is
also found for silicon (Fig.~\ref{fig:si_bands}). Therefore, a further
increase of the energy window is not very helpful anymore, although even
more compact virtual Wannier functions can be generated this way. 

There exist a couple of further interesting features in the disentangled
conduction bands of silicon. One is, that the second disentangled
band along the $\Sigma$ line climbs up to the {\it upper} part of the avoided
crossing in the middle of the $\Sigma$ panel. Apparently, close to this crossing
the contaminating $4s/4p$ orbital character is solely sitting on the 
energetically more {\it stable} second Bloch wave while the corresponding
valence $3s/3p$ contributions form the fourth BW, an interpretation which
is corroborated by a detailed analysis of the involved BWs.

The second point is the unexpected discontinuity in the uppermost disentangled
band along the $\Sigma$ line discernible in Fig.~\ref{fig:si_bands}a for the
$0-15$~eV window.
Its position coincides with the position of the avoided
crossing between the sixth and eighth conduction band which shows up in the
middle of the $\Sigma$ panel between 15 and 18~eV.
During the band disentanglement, the BWs from the upper part of this crossing
are excluded from the active space while those of the lower part are present,
and it seems that there is still a substantial part of the necessary orbital
character present in this upper BWs to cause the abrupt change in the
uppermost disentangled band. Obviously, the
$3s/3p$ valence orbital character moves up much higher into the unoccupied
band structure of silicon than one might expect at first glance. 

This is the reason why we switched to the larger energy window of $0-31$~eV, 
though a non-negligible upwards shift of the disentangled virtual bands arises.
Nevertheless we consider the disentangled bands and Bloch orbitals from the 
enlarged energy window to be the more appropriate ones.

\section{Conclusions}
\label{sec:concl}%
An extension of the Wannier-Boys localization algorithm for periodic
systems\cite{Baranek01,Zicovich01} is developed which allows to generate
localized Wannier functions in the case of {\it entangled} energy bands.
The method has been implemented into the localization
routine of the {\sc Crystal} program package.\cite{CRYSTAL00} Its main 
feature is the use of an enlarged set of active Bloch waves during the
optimization of the unitary hybridization matrix for the multi-band Wannier
transformation. This allows the inclusion of {\it all} Bloch waves which
contain noticeable admixture from orbitals with the same chemical character as
the localized Wannier functions one is looking for. The proper identification
of these admixtures is done by a simple projection technique.

The efficiency of our projective Wannier-Boys algorithm is demonstrated
for the virtual bands of three different systems,
{\it trans}-polyacetylene, diamond and bulk silicon. Localized {\it virtual}
Wannier functions could be generated in all three cases. The spatial extent
of them is found to be controllable by the size of the active space, i.e.,
the number of selected Bloch waves per $k$-point. The more Bloch waves
are considered, the more compact the localized Wannier functions become.
Yet, at the same time, an increasing tendency for an overall upwards shift
in the energies of the disentangled bands is observed. 

The same trend is
discernible in the Kohn-Sham energies of silicon discussed by Souza
{\it et al}.,~\cite{Souza01} though, because of the tight energy window
employed there (up to $\sim$11~eV), the effect is not very pronounced. The 
spread of the maximally localized virtual Wannier function of silicon
reported there is 30.13~bohr$^2$ (based on a $10\times 10\times 10$
Monkhorst-Pack grid) which is quite close to the value of 33.03~bohr$^2$
we found for the $0-15$~eV energy window.

The choice of the systems considered here was not accidental. Subsequent
use of the localized Wannier functions in wave-function-based post
Hartree-Fock correlation methods for periodic systems which explicitly
exploit the local character of virtual Wannier functions was the driving
force of this work, for example the method used in our study of the valence
and conduction bands of {\it trans}-polyacetylene\cite{Bezugly04} or in analogue
investigations of the band structure of bulk materials like
diamond.\cite{Willnauer04} The more compact the Wannier functions are
the better these local correlation methods perform.

\section{Acknowledgements}
We thank the {\sc Crystal} group in Torino for making available to us the source
code of the {\sc Crystal} 200x code without which the present study would not
have been possible.


\end{document}